\documentclass[aps,preprint,amsmath,amssymb]{revtex4-1}
\usepackage{graphicx}
\newcommand{\nn}{\nonumber}
\newcommand{\bd}{\begin{document}}
\newcommand{\ed}{\end{document}}
\newcommand{\bc}{\begin{center}}
\newcommand{\ec}{\end{center}}
\newcommand{\be}{\begin{eqnarray}}
\newcommand{\ee}{\end{eqnarray}}
\newcommand{\ba}{\begin{array}}
\newcommand{\ea}{\ed{array}}
\newcommand{\strich}[1]{#1  \! \! \slash}
\newcommand{\eqn}{\global\def\theequation}
\newcommand{\sw}{sin^2 \theta_W}
\newcommand{\fbd}{f_B}
\renewcommand{\thefootnote}{\alph{footnote}}
\newcommand{\se}{\section}
\newcommand{\sse}{\subsection}
\newcommand{\bi}{\bibitem}
\def\figcap{\section*{Figure Captions\markboth
     {FIGURECAPTIONS}{FIGURECAPTIONS}}\list
     {Figure \arabic{enumi}:\hfill}{\settowidth\labelwidth{Figure 999:}
     \leftmargin\labelwidth
     \advance\leftmargin\labelsep\usecounter{enumi}}}
\let\endfigcap\endlist \relax
\def\reflist{\section*{References\markboth
     {REFLIST}{REFLIST}}\list
     {[\arabic{enumi}]\hfill}{\settowidth\labelwidth{[999]}
     \leftmargin\labelwidth
     \advance\leftmargin\labelsep\usecounter{enumi}}}
\let\endreflist\endlist \relax

\begin{document}
\title
{\Large {\bf Study of the  Radivative  Pion Decay}}
% in the Light Front Quark Model}}
%form factors}

\author{ \bf \large Chuan-Hung Chen$^{1}$, Chao-Qiang Geng$^{2}$
  and Chong-Chung Lih$^{3}$\\
 }

\affiliation{  $^{1}$Department of Physics, National Cheng-Kung
University, Tainan 701, Taiwan \\
$^{2}$Department of Physics, National Tsing-Hua University,
Hsinchu
300, Taiwan  \\
$^3$Department of Optometry, Shu-Zen College of Medicine and Management,
Kaohsiung Hsien 452,Taiwan
 }

\date{\today}% It is always \today, today,

\begin{abstract}

We study the radiative pion decay of $\pi ^{+} \to e^{+}\nu_{e}\gamma$
in the  light front quark model (LFQM). We also summarize the result in 
the chiral perturbation theory.
The  vector and axial-vector hadronic form factors ($F_{V,A}$) for the
$\pi\to\gamma$ transition are evaluated in the whole allowed momentum transfer.
In terms of these momentum dependent form factors, we calculate the decay branching ratio and compare
our results with the experimental data and other theoretical predictions in the literature. We also
constrain the possible size of the tensor interaction in the LFQM.

\end{abstract}

\maketitle %

\se{Introduction}

The light pesudoscalar decays have been playing  important roles of understanding the standard model (SM).
In particular, the radiative
 pion decay of $\pi ^{+} \to e^{+}\nu_{e}\gamma$ ($\pi_{e2\gamma}$) is an interesting process, which can be used
 to test the $V-A$ structure of the weak interaction and search for  some
 anomalous interactions beyond the SM.
The decay consists of two types of contributions,
referred as internal-bremsstrahlung (IB) and
structure-dependent (SD) in terms of the emission of the photons, respectively.
The IB contribution  to the decay amplitude ($M_{IB}$)
is helicity suppressed like the $\pi_{e2}$ decay as
the photon radiates from the external electron,
while the SD one ($M_{SD}$), depending on vector and axial-vector weak hadronic currents,
is proportional to the electromagnetic coupling constant
$\alpha$ but free of the helicity suppression.
One can parametrize $M_{SD}$ by
the vector and axial-vector form factors, denote as $F_V$ and $F_A$, respectively.

The decay of $\pi ^{+} \to e^{+}\nu_{e}\gamma$ has been measured with the
branching ratio of (1.61$\pm$ 0.23)$\times 10^{-7}$
for the cuts of
$E_{\gamma}>$ 21 MeV and $E_{e}>$ $70-0.8 E_{\gamma}$
 by the ISTRA experiment \cite{istra,pdg}.
Recently, a more precise measurement on the decay branching ratio has been given by the PIBETA Collaboration \cite{pibeta1,pibeta2},
%i
with the decay branching ratios in various kinematic regions.
In particular, for the cuts of
$E_{e}>0.5$ MeV and $E_{\gamma}>10$ MeV
with the relative angle $\theta_{e\gamma}> 40^{ 0} $,  the decay branching ratio is
 $(73.86\pm0.54)\times 10^{-8}$~\cite{pibeta2}.
The new ongoing PEN experiment at PSI will at least  double the PIBETA data set \cite{PEN},
resulting in further improvements in precision \cite{Pocanic}.
In addition,
there is another  ongoing new experiment, PIENU, at TRIUMF \cite{TRIUMF} with a similar
sensitivity as the PEN experiment.

Theoretical calculations on $F_{V,A}$ as well as the decay branching ratio in the SM
have been done  in various QCD models \cite{tensor,tensor1,tensor2,ffdef,tensor3,pi4,largenc}.
In particular,
the decay branching ratio with
the same cuts
as those by ISTRA~\cite{istra,pdg} and PIBETA~\cite{pibeta2}
 is found to be  2.55$\times 10^{-7}$
 and 76.66$\times 10^{-8}$
in the chiral perturbation theory (ChPT) at $O(p^6)$
%,
\cite{mod2,mod3,mod4},  which are  larger the data shown above, respectively.
As a result, it may be necessary to consider some new types of interactions, such as
tensor interactions \cite{istra,tensor,tensor1,tensor2,ffdef,tensor3}.
It is clear that these tensor interactions are undoubtedly  signals
of new physics.
On the other hand, it is  important if we can obtain information on $F_{V,A}$ in some QCD models
other than the ChPT. For this purpose, in this study we will evaluate
$F_{V,A}$ in the light front quark model (LFQM)~\cite{LFQM,vex1}.
We will use the form factors in both  ChPT and
LFQM to examine the decay of $\pi^{+}\to e^{+}\nu_{e}\gamma$.
In addition, we will examine the new physics effect
due to the tensor interactions.

This paper is organized as follows. In Sec.~II,
we summarize the form factors in the $\pi \to \gamma$ transition
within the ChPT and LFQM.
In Sec.~III, we calculate the decay branching ratio of $\pi^+\to e^+\nu_e\gamma$ in these models.
We also compare our results with
 the experimental data and other theoretical predictions in the literature.
We give our conclusions in Sec.~IV.

\se{The form factors}
\subsection{Vector and Axial-vector Form Factors}
The decay amplitude for
$\pi ^{+} \to e^{+}\nu_{e}\gamma$ can be written
as:~\cite{Bryman,Chen1}
\be
M
%(\pi ^{+} \to e^{+}\nu_{e}\gamma)
&=&M_{IB}+M_{SD}\,,
\nn\\
M_{IB} &=&i{e \frac{G_{F}}{\sqrt{2}}}V_{ud}f_{\pi}m_e\epsilon^{*}_{\mu
}\,\bar{u}(p_{e})(1-\gamma _{5}) \left(\frac{p^{\mu }_{\pi}}{p_{\pi}\cdot q}-%
\frac{2p_{e}^{\mu }+\not{\! }q\gamma ^{\mu }}{2p_{e}\cdot q}%
 \right)v(p_{\nu})\,,
\nn\\
M_{SD} &=&-i{\frac{G_{F}}{\sqrt{2}}}V_{ud}\epsilon^{*}_{\mu }
\bar{u}(p_{e})\gamma _{\alpha }(1-\gamma
_{5})v(p_{\nu})
%\nn\\
%&&\times
\left[e\frac{F_{A}}{m_{\pi}}(-g^{\mu \alpha }p_{\pi}\cdot
q+p_{\pi}^{\mu }q^{\alpha })+ie\frac{F_{V}}{m_{\pi}}\epsilon ^{\mu \alpha \beta
\lambda }q_{\beta }p_{\pi \lambda }\right],\ \,
\label{n4}
\ee
where
$\epsilon_{\alpha}$ is the photon polarization vector,
$p_{\pi}$, $p_{e}$, $p_{\nu}$, and $q$
are the four momenta of $\pi^{+}$, $e^{+}$,
$\nu$ and $\gamma$,
and  $f_{\pi}$ and $F_{A,V}$
are the $\pi$ meson decay constant
and  the axial-vector and vector form factors,
 defined by
\be
\langle\, 0|\bar{s}\gamma^{\mu}\gamma_5 u|\pi^+(p_{\pi}) \,\rangle &=&if_{\pi}p_{\pi}^{\mu},
\nn\\
\langle\gamma(q) |\bar{u}\gamma_{\mu }\gamma _{5}d|\pi(p_{\pi}) \,\rangle &=&
e{\frac{F_{A}}{%
m_{\pi}}}\left[ (p\cdot q) \epsilon ^*_\mu
-(\epsilon ^{*}\cdot p)q_{\mu }\right] ,
\nonumber\\
\langle\gamma(q) |\bar{u}\gamma_{\mu }d|\pi(p_{\pi}) \,\rangle &=&
ie{\frac{F_{V}}{m_{\pi}}}\varepsilon^{\mu \alpha \beta \nu }
\epsilon^*_{\alpha }q_{\beta }p_\nu \, ,
\label{4}
\ee
respectively,
with $p=p_{\pi} -q$ being the transfer momentum.
Obviously,  $M_{IB}$ has a suppression factor of $m_{e}$.
The physically accessible kinematics
region is $0\leq p^{2}\leq p_{\max }^{2}=m_{\pi}^{2}$ due to the time-like
momentum transfers.
In the following discussion, we will
 first
summarize the formulas for $F_{V,A}$ in the ChPT \cite{mod3,mod4}
and
then  evaluate these form factors
in the
LFQM. We note that similar calculations for the
$P\to \gamma\ (P=K^+,K^0,D,B)$ transitions in the LFQM have
been performed in Refs. \cite{lf1,mod1,lf2}.

\subsubsection{Chiral Perturbation Theory}

The tree and loop contributions to $F_{V,A}$ in the ChPT at $O(p^6)$
for the $\pi_{e 2\gamma}$ decay have been calculated in Refs. \cite{mod3,mod4}.
The explicit forms can be summarized as \cite{lf1}
\be
F_{V}(p^2)&=&\frac{m_{\pi}}{4\sqrt{2}\,\pi^{2}F_{\pi}}
\bigg\{1-\frac{256}{3}\pi^{2}m_{K}^{2}C_{7}^{r}
+\frac{64}{3}\pi^{2} p^{2} C_{22}^{r}
\nonumber\\
&-&\frac{1}{8 \pi^{2} F_{\pi}^{2}}\bigg[m_{\pi}^{2}
\ln \left(\frac{m_{\pi}^{2}}{\mu^{2}}\right)
+m_{K}^{2} \ln \left(\frac{m_{K}^{2}}{\mu^{2}}\right)
\nonumber\\
&-&\int \left[m_{K}^{2}-x(1-x)p^{2}\right]
\ln\left(\frac{m_{K}^{2}-x(1-x)p^{2}}{\mu^{2}}\right)dx
\nonumber\\
&-&\int \left[m_{\pi}^{2}-x(1-x)p^{2}\right]
\ln\left(\frac{m_{\pi}^{2}-x(1-x)p^{2}}{\mu^{2}}\right)dx
\bigg]\bigg\}\,,
\label{fvk}
\ee
and
\be
F_{A}(p^2)&=&\frac{4 \sqrt{2}\,m_{\pi}}{F_{\pi}}(L _{9}^{r}+L _{10}^{r})-
\frac{m_{\pi}}{6F_{\pi}^{3}(2\pi)^{8}}[22.25\,(m_{\pi}^{2}-p^{2})+193.4]
\nonumber \\
&-&\frac{m_{\pi}}{2\sqrt{2}\,\pi^{2}F_{\pi}^{3}}\bigg\{ (L
_{3}^{r}+2L _{9}^{r}+2L _{10}^{r})\,m_{K}^{2}\ln \Big(\frac{m_{K}^{2}}{m_{\rho }^{2}}\Big)
\nonumber \\
&+&2\,(2L_{1}^{r}-L _{2}^{r}+L _{3}^{r}+2L _{9}^{r}
+2L _{10}^{r})m_{\pi}^{2}\ln \Big(\frac{m_{\pi}^{2}}{m_{\rho }^{2}}\Big)\bigg\}
\nonumber \\
&-&\frac{4\sqrt{2}\,m_{\pi}}{F_{\pi}^{3}}\bigg\{
4m_{K}^{2}(6y_{18}^{r}-2y_{82}^{r}+y_{84}^{r}+2y_{103}^{r}) \nonumber \\
&+&2m_{\pi}^{2}(6y_{17}^{r}+6y_{18}^{r}-2y_{81}^{r}
-2y_{82}^{r}+2y_{83}^{r}+y_{84}^{r}+y_{85}^{r}-y_{100}^{r}+2y_{102}^{r}
\nonumber \\
&+&2y_{103}^{r}-2y_{104}^{r}+y_{109}^{r})+\frac{1}{2}(m_{\pi}^{2}-p^{2})(2y_{100}^{r}
-4y_{109}^{r}+y_{110}^{r})\bigg\} \,,
\label{fak}
\ee
where the wave function and decay constant ($F_\pi\equiv f_\pi/\sqrt{2}$) renormalizations have been included and
$C^{r}_{i}$, $L^{r}_{i}$ and $y^{r}_{i}$
are the renormalized coupling constants.
  Note that the first terms in Eqs. (\ref{fvk})
and (\ref{fak}) correspond to $F_V$ and $F_A$ at $O(p^4)$
\cite{mod2,kl2m},  respectively.
To get the numerical results for the form factors, we take
$m_K=0.495$ GeV, $m_\pi=0.14$ GeV and $m_\rho=0.77$ GeV, $F_{\pi}=0.092$ GeV
and the renormalized coefficients of
$(L^r_1,L^r_2,L^r_3,L^r_9,L^r_{10})$, $(C^r_7,C^r_{22})$ and
$(y^r_{100},y^r_{104},y^r_{109},y^r_{110})$ to be
$(0.53,0.71,-2.72,6.9,-5.5)\times 10^{-3}$ \cite{33},
$(0.013,6.52)\times 10^{-3}\,GeV^{-2}$ \cite{29}
and $(1.09,-0.36,0.40,-0.52)\times 10^{-4}/F_{\pi}^2$ \cite{25}, respectively.
For some other possible sets of coefficients, see Ref. \cite{mod4}
as well as the recent review in Ref. \cite{Bi2007}.
Note that the uncertainties for the renormalized coupling constants are not considered in this study.

\subsubsection{Light Front Quark Model}

In the light front (LF) approach, the general structure of the phenomenological
LF meson wave function is based only on the $Q\bar{q}$
Fock space sector \cite{lf1}. The pion wave function can be expressed
by an anti-quark $\bar{q}$
and a quark $Q$ with the total momentum $(p+q)$  as:
\begin{eqnarray}
|\pi (p+q)\,\rangle&=& \sum_{\lambda _{1}\lambda_{2}}\int [dk_{1}][dk_{2}]
2(2\pi)^{3}\delta ^{3}(p+q-k_{1}-k_{2})  \nonumber \\
&& ~~~~~~~~ \times \Phi _{\pi}^{\lambda _{1}\lambda _{2}}(z,k_{\bot})
b_{\bar{q}}^{+}(k_{1},\lambda _{1}) d_{Q}^{+}( k_{2},\lambda _{2})
|0\,\rangle\,,
\end{eqnarray}
where $\Phi _{\pi}^{\lambda _{1}\lambda _{2}}$ is the amplitude of the corresponding
$\bar{q}(Q)$ and
$k_{1(2)}$ is the on-mass shell LF momentum of the internal
quark. The LF relative momentum variables $(z,k_{\bot})$ are defined by
\begin{eqnarray}
&& k^+_1=(1-z) (p+q)^+, \quad k^+_2=z (p+q)^+\,,  \nonumber \\
&& k_{1\bot}=(1-z) (p+q)_\bot+k_\bot, \quad k_{2\bot}=z
(p+q)_\bot-k_\bot\,,
\end{eqnarray}
and
\be
\Phi _{\pi}^{\lambda _{1}\lambda _{2}}(z,k_{\bot })&=&\left( \frac{%
k_{1}^{+}k_{2}^{+}}{2[M_{0}^{2}-\left( m_{Q}-m_{\bar{q}} \right) ^{2}]}\right)^{%
\frac{1}{2}}\overline{u}\left( k_{1}, \lambda _{1}\right)
\gamma^{5}v\left( k_{2},\lambda _{2}\right) \phi(z,k_{\bot}) \,,
\nn\\
M_0^2&=&{\frac{k^2_{\bot}+m_q^2}{1-z}}+{\frac{k^2_{\bot}+m_Q^2}{%
z}} \, . \label{n6} \ee where $\phi(z,k_{\bot})$ is the space part
of the wave function,
%,
which is taken to be a Gaussian type but it
can be solved in principle by
the LF QCD bound state equation \cite{lf2}.
At the quark loop diagram, the hadronic matrix elements in
Eq. (\ref{4})
can be obtained to be \be &&\langle\gamma (q)|\bar{u}\gamma_{\mu
}\,(1-\gamma _{5})\,d|\,\pi(p+q)\, \rangle=\int \frac{d^{4}k_1}{(2
\pi)^{4}} \Lambda_{\pi} \nonumber \\ &&\times\bigg\{
\gamma_{5}\frac{i(-\strich{k}_{2}+m_{u})}
{k_{2}^{2}-m_{u}^{2}+i\epsilon} ie_{u}\strich{\epsilon}^*
\frac{i(\strich{k}_{3}+m_{u})} {k_{3}^{2}-m_{u}^{2}+i\epsilon}
\gamma_{\mu }(1-\gamma _{5})\frac{i(\strich{k}_{1}+m_{d})}
{k_{1}^{2}-m_{d}^{2}+i\epsilon} \nonumber \\ &&+(u\leftrightarrow
d \, , k_{1} \leftrightarrow k_{2} )\bigg\}\,,
\label{int}
\ee
where $\Lambda_{\pi}$ is a vertex function related to the
quark-antiquark bound state of the $\pi$ meson, $k_2=q-k_3$ and
$k_1=(p+q)-k_2=k_3+p$. By integrating over the LF momentum
$k_{2}^{-}$ in Eq. (\ref{int}), we get
\be
&&\langle\gamma
(q)|\bar{u}\gamma_{\mu }\,(1-\gamma
_{5})\,d|\,\pi(p+q)\, \rangle\nonumber \\
&& =\int_{p}^{p+q}
[d^{3}k_{1}]\bigg\{\frac{\Lambda_{\pi}}{k_{1}^{-}-k_{1on}^{-}}
(I^{\mu}|_{k_{2on}^{-}}) \frac{1}{k_{3}^{-}-k_{3on}^{-}}
+(u\leftrightarrow d \, , k_{1} \leftrightarrow k_{2})\bigg\}
\,\,, \label{22} \ee where \be
&&[d^{3}k_{1}]=\frac{dk_{1}^{+}dk_{1\bot}}{2(2\pi)^{3} k_1^{+}
k_2^{+} k_3^{+}} ~, \nonumber \\
&&I^{\mu}|_{k_{2on}^{-}}=Tr\bigg\{
\gamma_{5}(-\strich{k}_{2}+m_{u})ie_{q}\strich{\epsilon}^*
(\strich{k}_{3}+m_{u})\gamma_{\mu } (1-\gamma
_{5})(\strich{k}_{1}+m_{d})\bigg\}~, \nonumber \\
&&k_{ion}^{-}=\frac{m_{i}^{2}+k_{i\bot}^{2}}{k_{i}^{+}}~,~
k_{1(2)}^{-}=p_{on}^{-}-k_{2(1)on}^{-} ~,~
k_{3}^{-}=q^{-}-k_{1on}^{-} \,,
\label{trace}%
\ee
with $\{on\}$ representing the on-shell particles.
Note that the vertex function $\Lambda_{\pi}$ in Eqs. (\ref{int})
and (\ref{22}) include the normalization factor of the wave
function and momentum  distribution function, given by \cite{vex1}:
\be \frac{\Lambda_{\pi}}{k_{1}^{-}-k_{1on}^{-}} = {\sqrt{k_1^{+}
k_2^{+}}\over \sqrt{2} ~{ M_0}}~ \phi(z, k_\bot)\,.
\label{E11}
\ee %
Note that in Eq. (\ref{E11}), we have take $m_q=m_Q$, i.e., $m_u=m_d$ for $\pi$.
To calculate the matrix element in
Eq. (\ref{22}), we choose
a frame with the transverse momentum $p_{\bot}$ = $0$ so that
$p^{2}=p^{+}p^{-} \geq 0$ covers the entire range of the momentum
transfers.
Here,
the relevant quark momentum variables are
\be
k_{3}^{+}=(1-z')q^+,~~k_{2}^{+}=z' q^+,
~~k_{3\perp}=(1-z')q_\perp+k'_\perp,~~k_{2\perp}=z'
q_\perp-k'_\perp\,.
\label{transmom}
\ee
By considering the good
component as ``$\mu=+$'', the 
%last two
hadronic matrix elements in
Eq. (\ref{4}) can be rewritten as: 
\be 
\langle\, 0|\bar{s}\gamma^{+}\gamma_5 u|\pi(p+q) \,\rangle &=&if_{\pi}(p+q)^{+}\,,
                \nonumber \\
\langle\gamma (q)|
\bar{u}\gamma^{+}\gamma _{5}d|\pi(p+q)\,\rangle
&=&-e\frac{F_{A}}{2m_{\pi}}\left( \epsilon
        _{\bot }^{*}\cdot q_{\bot }\right) p^{+}\,,
                \nonumber \\
\langle\gamma (q)| \bar{u}\gamma^{+}d|\pi(p+q)\,\rangle
&=&-ie\frac{F_{V}}{2m_{\pi}}
\epsilon ^{ij}\epsilon _{i}^{*}q_{j}p^{+}\,.
\label{ff}
\ee
Using  Eq.~(\ref{transmom}), the trace part $I^{\mu}$ in
Eq.~(\ref{trace}) can be carried out.
By comparing  the last two equations in Eq.~(\ref{ff}) with  those in Eq.~(\ref{22}), we derive
\be
F_{A}(p^2) &=&4m_{\pi}
        \int \frac{dz'\,d^{2}k_{\bot }}{2(2\pi)^{3}}\Phi
        \left( z,k_{\bot }^{2}\right) {1\over 1-z}
 %               \nonumber \\
%&&~~~~~~~~ \times
\left\{ \frac{1}{3}\frac{m_{d}+Bk_{\bot }^{2}
        \Theta}{m_{d}^{2}+
        k_{\bot}^{2}}+\frac{2}{3}\frac{m_{u}-Ak_{\bot }^{2}\Theta }
        {m_{u}^{2}+k_{\bot }^{2}}  \right\}\,,   \nonumber \\
F_{V}(p^2) &=&-4m_{\pi}
        \int \frac{dz'\,d^{2}k_{\bot }}{2\left( 2\pi \right) ^{3}}\Phi
        \left( z,k_{\bot }^{2}\right) {1\over 1-z}
                \nonumber \\
&&\times\left\{ \frac{1}{3}\frac{m_{d}+(1-z)(m_{d}-m_{u}) k_{\bot
}^{2}
        \Theta }{m_{d}^{2}+k_{\bot }^{2}}-\frac{2}{3}\frac{m_{u}-
        z\left( m_{d}-m_{u}\right) k_{\bot }^{2}\Theta }{m_{u}^{2}
        +k_{\bot }^{2}}\right\}\,,
\label{fav}
\ee
where
\be
A &=& z(1-2z') (m_d-m_u) -2 z' m_u\,, \nn\\
B &=& z(1-2z') m_d+m_d+(1-2z') (1-z)m_u\,,  \nn\\
\Phi (z,k_{\bot}^2) &=&
4 \left({\frac{\pi}{\omega_{\pi}^{2}}}\right)^{\frac{3}{4}}
\left( {\frac{z(1-z) }{2[M_0^2-(m_d-m_u)^2]}}%
\right)^{1/2} \sqrt{{\frac{dk_{z}}{dz}}}\exp \left( -{\frac{\vec{k}^{2}}{%
2\omega_{\pi}^2}}\right)\,,  \nonumber \\
\Theta &=& {\frac{1}{\Phi(z,k_{\bot}^2) }} {\frac{d\Phi(z,k_{\bot}^{2})}{%
dk_{\bot}^2}} \, ,  \nonumber \\
z&=&z^{\prime}\left(1-{\frac{p^2}{m_{\pi}^2}}\right),\
\vec{k}=(\vec{k}_{\bot}, \vec{k}_{z}) \,,
%\ee
%\be
\nonumber\\
k_{z}&=& \left( z-\frac{1}{2}\right) M_{0}+\frac{m_{d}^{2}-m_{u}^{2}}{%
2M_{0}} \, .
\ee
We note that to evaluate the form factors, we have to fix the meson scale parameter $\omega_{\pi}$
in the meson wave functions by fitting the meson decay constant, given by~\cite{lfpa}
% which can be calculated by Eq.~(5-7)\cite{lfpa}
\be
f_{\pi}&=&\,{\sqrt{48}}\int {dz\,d^2k_\perp\over 2(2\pi)^3}\,\Phi(z,
k_\perp)\,\frac{m_{u}}{z(1-z)}\,.
\label{fpi}
\ee

\sse{Tensor Form Factor}

The tensor interaction is
 given by \cite{tensor,ffdef}
\be
{\cal L}_{T}&=&\frac{G_{F}}{\sqrt{2}}\sin\theta_{c}  f_{T}
\left(\,\bar{u}\,\sigma_{\mu\nu}\gamma_{5}\,d\,\right)
\left[\,\bar{e}\,\sigma^{\mu\nu}(1-\gamma_{5})\,\nu_{e}\,\right].
\label{tensorl}
\ee
The tensor form factor is defined by \cite{ffdef}:
\be
\langle\gamma(q) |\bar{u}\sigma_{\mu\nu }\gamma_{5}d|\pi(p_{\pi}) \,\rangle &=&
-i\frac{eF_{T}}{2f_{T}}\left(\,\epsilon^*_{\mu} q_{\nu}
- q_{\mu } \epsilon^*_{\nu}\,\right)\,.
\label{tff4}
\ee
For
the LF good component of ``$\mu=+$'', one rewrites Eq. (\ref{tff4}) as
\be
\langle\gamma (q)|\bar{u}\sigma^{+\nu }\gamma _{5}d|\pi(p+q)\,\rangle
&=&-i\frac{eF_{T}}{2f_{T}} \left( \epsilon_{\bot }^{*}\cdot q_{\bot }\right)\,.
\label{tff}
\ee
At the quark level, the hadronic matrix element in
Eq. (\ref{tff4})
%,
is found to be
\be
\langle\gamma (q)|\bar{u}\sigma_{\mu\nu }\gamma_{5}\,
d|\,\pi(p+q)\, \rangle=~~~~~~~~~~~~~~~~~~~~~~~~~~~~~~~~~~~~~~~~~~~~~~~~~~~~~~~~~~~~~~~~~~~~~~
\nonumber \\
\int \frac{d^{4}k_1'}{(2 \pi)^{4}}
\Lambda_{\pi}
 %\nonumber \\ &&\times
 \bigg\{
\gamma_{5}\frac{i(-\strich{k}_{2}+m_{u})}
{k_{2}^{2}-m_{u}^{2}+i\epsilon} ie_{u}\strich{\epsilon}^*
\frac{i(\strich{k}_{3}+m_{u})} {k_{3}^{2}-m_{u}^{2}+i\epsilon}
\sigma_{\mu\nu }\gamma_{5}\frac{i(\strich{k}_{1}+m_{d})}
{k_{1}^{2}-m_{d}^{2}+i\epsilon}
%\nonumber \\
%&&
+(u\leftrightarrow d , k_{1} \leftrightarrow k_{2}
)\bigg\},~~~~
%~~~~~~~~~~~~~~~~~~~~~~~~~~~~~~~~~~~~
\label{tint}
\ee
which leads to
\be
&&\langle\gamma (q)|\bar{u}\sigma_{\mu }\,\gamma_{5}\,d|\,\pi(p+q)\, \rangle
=
\nonumber\\
&&\int_{p}^{p+q}
[d^{3}k_{1}]\bigg\{\frac{\Lambda_{\pi}}{k_{1}^{-}-k_{1on}^{-}}
(I^{\mu\nu}|_{k_{2on}^{-}}) \frac{1}{k_{3}^{-}-k_{3on}^{-}}
+(u\leftrightarrow d \, , k_{1} \leftrightarrow k_{2})\bigg\}\,,
%\nonumber \\
\label{t22}
\ee
where
\be
I^{\mu\nu}|_{k_{2on}^{-}}=Tr\bigg\{
\gamma_{5}(-\strich{k}_{2}+m_{u})ie_{q}\strich{\epsilon}^*
(\strich{k}_{3}+m_{u})\sigma_{\mu } \gamma_{5}
(\strich{k}_{1}+m_{d})\bigg\}\,.
\ee
 From Eqs. (\ref{tff}) and (\ref{t22}),
we obtain
\be
\frac{F_{T}(p^2)}{f_{T}} &=&2 \int
\frac{dz'\,d^{2}k_{\bot }}{2\left( 2\pi \right) ^{3}}\Phi
        \left( z,k_{\bot }^{2}\right) \left\{
        \frac{2}{3}\frac{C_{1}+C_{2}k_{\bot }^{2}
        \Theta}{m_{u}^{2}+k_{\bot}^{2}}
        +\frac{1}{3}(m_{d}\leftrightarrow m_{u}) \right\} \,,
\label{ttff}
\ee
where
\be
C_1 &=&
\frac{1}{z'z(1-z)^{2}(1-z')}\Bigg\{
(1-2z'+2z'^{2}-z'z)(z'+z-2z'z)k_{\bot}^{2}  \nn \\
&&+(1-z)m_u \left[2z'z(1-z')m_d
+(z'+z+2z'^{2}z-4z'z)m_{d}\right]\Bigg\}\,, \nonumber \\
C_2 &=& \frac{(z-z')}{z'z(1-z)^{2}(1-z')}\Bigg\{
(z'+z-2z'z) k_{\bot}^{2}+z^{2}(1-z')m_{d}^{2}\nn \\
&& -(1-z)(z'+z-z'z)m_{u}^{2}\Bigg\}\,.
 \ee 
 At the maximal recoil
of $p^2=0$, we have \be \frac{F_{T}(0)}{f_{T}} &=&4 \int
\frac{dz\,d^{2}k_{\bot }}{2\left( 2\pi \right) ^{3}}\Phi \left(
z,k_{\bot }^{2}\right) \bigg\{
\frac{2}{3}\frac{(1-z)\,k^{2}+m_{u}\,(z\, m_{d}+(1-z)\, m_{u})}
{z\,(m_{u}^{2}+k_{\bot}^{2})} \nn \\
&&~~~~+\frac{1}{3}(m_{u}\leftrightarrow m_{d}) \bigg\}\, .
\label{ttff0}
\ee

\sse{Numerical results}

\begin{figure}[htbp]
\includegraphics*[width=4in]{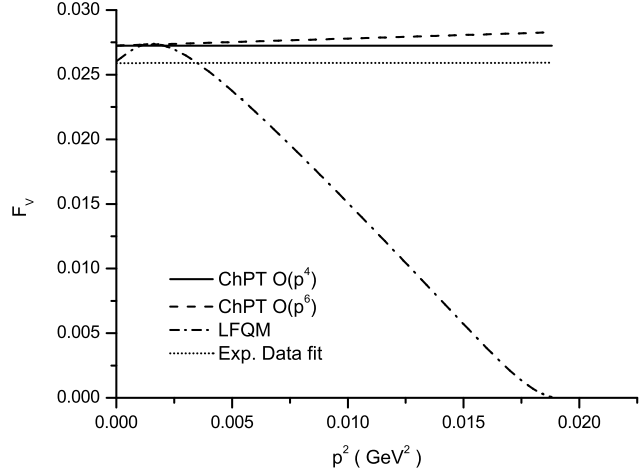}
\caption{ $F_V(p^2)$ as a
function of the transfer momentum $p^2$ with $m_u=m_d=250$~MeV. }
\label{FV}
\end{figure}

\begin{figure}[htbp]
\includegraphics*[width=4in]{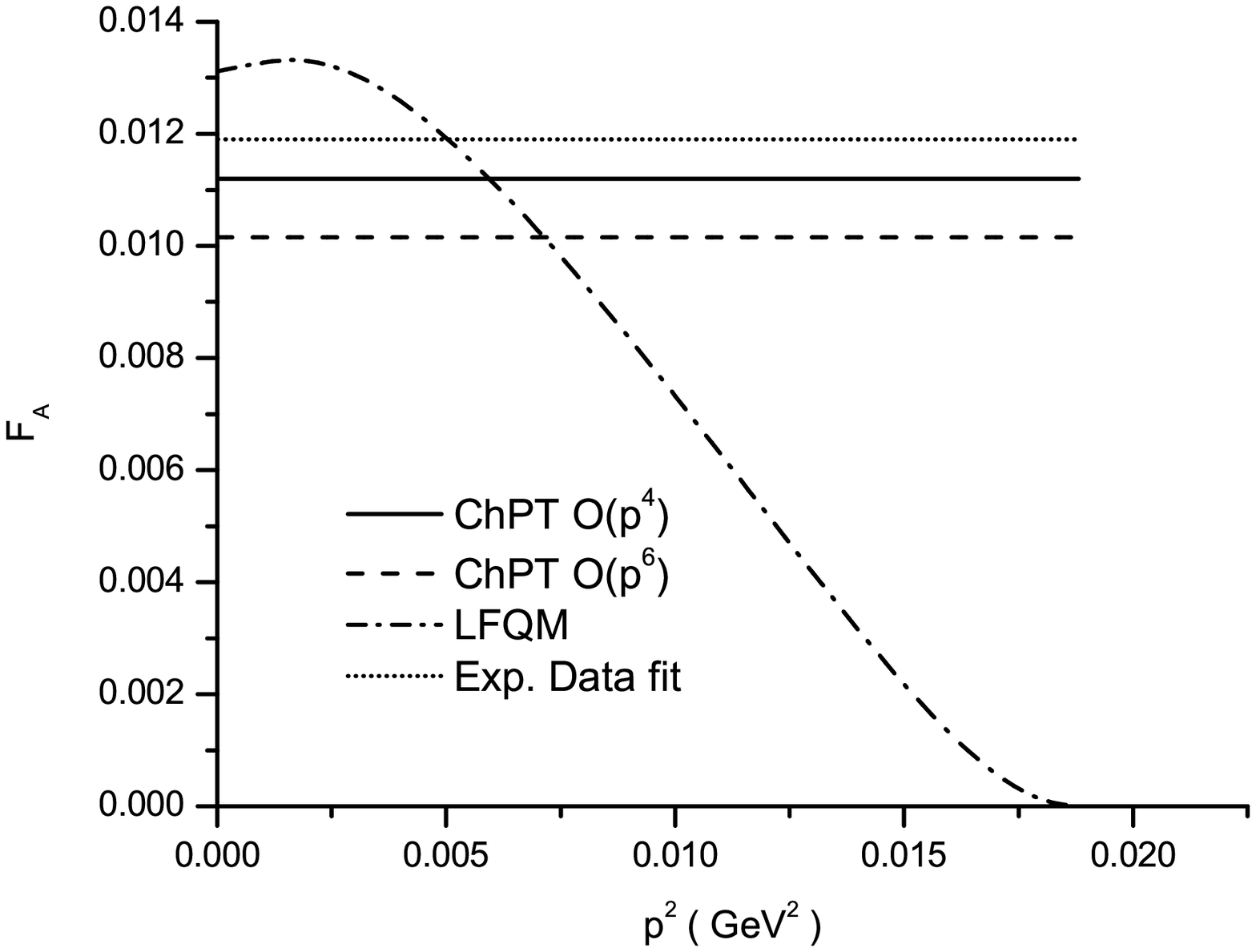}
\caption{ Same as Fig.~\ref{FV} but for $F_V(p^2)$.}
%$F_A(p^2)$ as a function of the transfer momentum $p^2$. }
\label{FA}
\end{figure}

\begin{figure}[htbp]
\includegraphics*[width=4in]{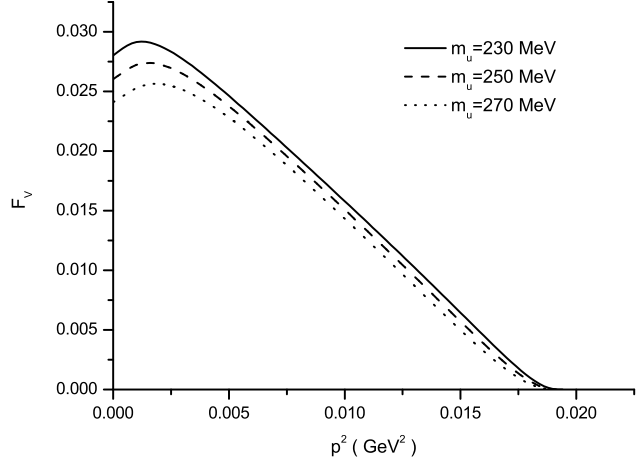}
\caption{ $F_V(p^2)$ in the LFQM, where $m_u=m_d=230$, $250$ and $270$~MeV correspond to
solid, long-dashed and short-dashed lines, respectively.}
% as a function of the transfer momentum $p^2$ with $m_u=m_d=250$~MeV. }
\label{FVqm}
\end{figure}

\begin{figure}[htbp]
\includegraphics*[width=4in]{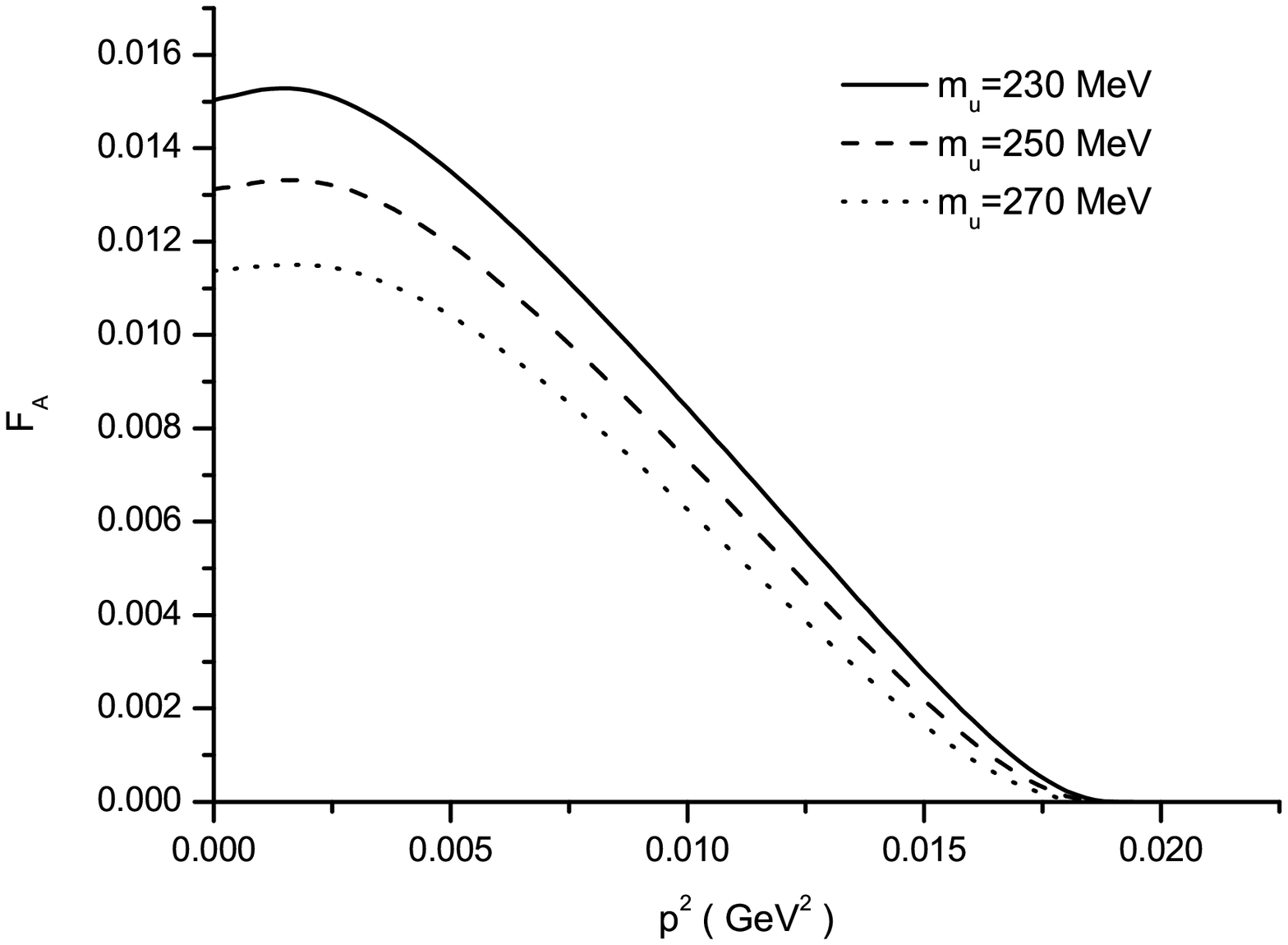}
\caption{ Same as Fig.~\ref{FVqm} but for $F_V(p^2)$.}
%$F_A(p^2)$ as a function of the transfer momentum $p^2$. }
\label{FAqm}
\end{figure}

To compute numerical values of the form factors in the LFQM,
the $\omega$ parameter in the light-front
wave function is fixed by other hadronic properties. 
For example, 
by using the decay
constant of $f_{\pi}=130$~MeV and the quark masses of $m_{u}=m_{d}=250$~MeV,
we obtain $\omega_{\pi}=301$~MeV from Eq.~(\ref{fpi}).
We note that this value of $\omega_{\pi}$ is just a typical one and its uncertainty 
%of $\omega_{\pi}$ 
mainly arises from those of the light quark masses.
%, which will not be discussed in this work. 
%%%%%%
The transfer momentum $p^2$ dependences
of $F_V$ and $F_A$ are shown in Figs.~\ref{FV} and~\ref{FA}, respectively.
Note that the behaviors of the figure's sharps are independent  of the quark masses but the 
values of $F_V$ and $F_A$ at  the maximal recoil of $p^2=0$ can be quite different as shown in Table~\ref{T2}.
The results in Table~\ref{T2} also illustrate the main uncertainty from the quark masses.
In Figs.~\ref{FV} and~\ref{FA}, we have also included the experimental results fitted by the forms of
$F_V(p^2)=F_V(1+\alpha p^2)$ and $F_A(p^2)=F_A(0)$ with a constant parameter of $\alpha$~\cite{pibeta2},
%At the maximal recoil of $p^2=0$,
while in Table.~\ref{T2}, we list the values of $F_{A,V}(0)$ in the ChPT and  those from the  data.
%are listed in Table.~\ref{T2}. 
We remark that the numerical values of the form factors at $p^2=0$ 
for the pion case between the theoretical models seem to be 
less compatible comparing with those for the kaon in Ref.~\cite{mod4} in which the strange quark mass also enters.
To illustrate the quark mass effects on the form factors in the LFQM, 
in Figs.~\ref{FVqm} and~\ref{FAqm} we plot three different sets of quark masses including the one 
in Figs.~\ref{FV} and \ref{FA}. It is clear that both $F_{V,A}$ decrease as $m_{u,d}$ increase.



\begin{table}[htbp]
\caption{Values of $F_{A,V}(0)$ in
(a) the ChPT at $O(p^4)$, (b) the ChPT at $O(p^6)$,
(ci) the LFQM with i=1, 2 and 3 for $m_{u,d}=230$, $250$ and $270$~MeV, repectively.}
\vskip 0.2in
\label{Table1}
\begin{tabular}{|c||c|c|c|c|c|c|c|} \hline
 & (a) & (b) & (c1) & (c2) & (c3) & Data \cite{pibeta2}
%& Data in \cite{pi2} & \cite{pi4}
\\ \hline \hline
$F_{A}(0)$ & $ 0.0112 $
& $ 0.0102 $ & $ 0.0151 $ & $ 0.0131 $ & $ 0.0113 $ & $0.0117(17) $
%& $0.0121(18) $  & $ 0.0121(18) $
\\ \hline
$F_{V}(0)$ &$ 0.0272 $
& $ 0.0272 $ & $ 0.02751 $ & $0.0261 $ & $ 0.0243 $ & $0.0258(17) $
%& $0.0258(18) $  & $ 0.0271 $
\\ \hline
\end{tabular}
\label{T2}
\end{table}
The tensor form factor Eq. (\ref{ttff})
in the whole kinematic region for the LFQM is  shown in Fig. 5.
\begin{figure}[htbp]
\includegraphics*[width=4in]{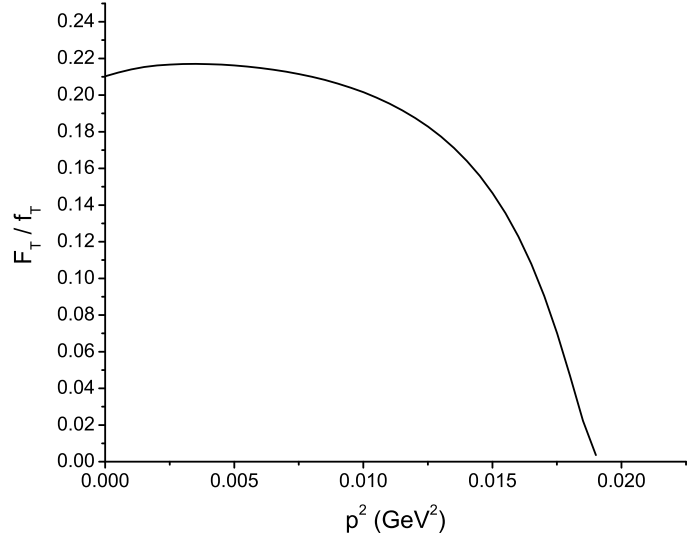}
\caption{ $\frac{F_T(p^2)}{f_{T}}$ as a
function of the transfer momentum $p^2$. }
\label{F3}
\end{figure}
At $p^2=0$, we get the $F_{T}(0) /f_{T}$ =$0.220$, $0.210$ and $0.202$ for $m_{u,d}=230$, $250$ and $270$~MeV,
respectively.

\se{Decay Branching Ratio}

In the $\pi^{+}$ rest frame, we obtain the double differential decay rate
as
\be
\frac{d^{2}\Gamma ^{l}}{dx\, dy }
&=&\frac{m_{\pi}}{256\pi
^{3}}\left| M\right| ^{2}=\frac{\alpha}{2\pi
} Br(\pi \to e \nu ) A,
\ee
%with
\be
A&=&A_{IB}(x,y)+A_{SD}(x,y)+A_{INT}(x,y)\,,
\label{43}
\ee
with
\be
A_{IB}(x,y) &=&
{\frac{1-\lambda}{\lambda x^{2}}}
\left[ x^{2}+2(1-r_e)
\left(1-x-{\frac{r_e}{\lambda}}\right)\right]\,,
  \nonumber \\
A_{SD}(x,y) &=&
\frac{m_{\pi}^{4}(1-\lambda)}{4 f^2_{\pi}m_{e}^{2}}x^{2}\left[ |F_{V}+F_{A}|^{2}
{\frac{\lambda ^{2}}{1-\lambda }}
\left( 1-x-{\frac{r_{e}}{\lambda }}\right) \right.   \nonumber \\
&&+\left. |F_{V}-F_{A}|^{2}(y-\lambda
)\right]\,,
  \nonumber \\
A_{IN}(x,y) &=&
-\frac{m_{\pi}}{f_{\pi}}\left[ Re[(F_{V}+F_{A})^{*}]\left( 1-x
-{\frac{
r_e}{\lambda }}\right) \right.   \nonumber \\
&&-\left. Re[(F_{V}-F_{A})^{*}]{\frac{1-y+\lambda }{\lambda }} \right] \,,
\label{48}
\ee
where $x=2E_{\gamma}/m_{\pi}$, $y=2E_e/m_{\pi}$, $r_e=m_e^2/m^2_{\pi}$ and
$\lambda=(x+y-1-r_e)/x$.
One can also relate the
angle $\theta_{e\gamma}$ between
the $e^+$ and photon momenta with $y$ and $\lambda$. Explicitly,
by neglecting the $r_e$, one has
that
%.
\be
\lambda=y \sin^{2} \left (\frac{\theta_{e\gamma}}{2} \right)\,.
\ee
The physical regions for $x$ and $y$ are given by
\be
0 \le &x& \le 1-r_e  \,, \nonumber \\
1-x+\frac{r_e}{1-x} \le &y& \le 1+r_e\,.
\ee
In Table II,  we show the decay branching fractions
of $\pi^{\pm} \to e^{\pm}\nu_{e}\gamma$ in terms of the various contributions in Eq. (\ref{48})
for
$f_{\pi}=130$ MeV, $m_{\pi}=140$  GeV \cite{pdg}, $m_{u,d}=250$~MeV
and $Br(\pi \to e \nu)=(1.23\pm 0.004)\times 10^{-8}$
with the cuts of $E_{\gamma} > $ 50 MeV and
$E_{e} > $ 50 MeV.
Note that in this kinematic region, the contribution from the SD part dominates the decay
rate, which is sensitive to the $V-A$ structure as well as  new physics.
We note that the total branching ratio in the LFQM are 2.937 and $2.320\times 10^{-8}$ for 
$m_{u,d}=230$ and $270$~MeV, respectively.
\begin{table}[htbp]
\caption{ Decay branching ratio  of $\pi \to e \nu_e \gamma$ (in units of $10^{-8}$)
in (a) the ChPT at $O(p^4)$, (b) the ChPT at $O(p^6)$
and (c) the LFQM with the cuts of $E_{\gamma} > $ 50 MeV and
$E_{e} > $ 50 MeV, respectively.}
 \vskip 0.2in
\label{Table2}
\begin{tabular}{|c||c|c|c|c|} \hline
Model & IB & SD & INT & Total
\\ \hline \hline
(a) & $ 3.692 \times 10^{-1} $
& $ 2.356 $ & $2.536\times 10^{-3} $ & $ 2.727 $
\\ \hline
(b) &$ 3.692 \times 10^{-1} $
& $ 2.309 $ & $2.850\times 10^{-3} $ & $ 2.679 $
\\ \hline
(c) &$ 3.692 \times 10^{-1} $
& $ 2.250 $ & $1.840\times 10^{-3} $ & $ 2.621 $
\\ \hline
\end{tabular}
\end{table}

\begin{table}[htbp]
\caption{Decay branching ratio  of $\pi \to e \nu_e \gamma$
(in units of $10^{-8}$) in
(a) the ChPT at $O(p^4)$, (b) the ChPT at $O(p^6)$,
(ci) the LFQM with i=1, 2 and 3 for $m_{u,d}=230$, $250$ and $270$~MeV, repectively,
%(c) the LFQM, 
(d) the green function method~\cite{pi4} and (e) the ChPT with a large $N_C$ expansion~\cite{largenc}
as well as  the data in Ref.~\cite{pibeta2} in various kinematic energy regions (in units of MeV).
 }
 \vskip 0.2in
 \label{Table3}
\begin{tabular}{|c|c|c||c|c|c|c|c|c|c|c|} \hline
$E_{e}^{min}$& $E_{\gamma}^{min}$ & $\theta_{e\gamma}^{min}$ &
Data~\cite{pibeta2} &
 (a) & (b) & (c1) & (c2) &(c3) &(d) & (e)
\\ \hline \hline
$50$ & $50$ &   $ -$       & $2.614(21)$ & $2.727(9)$   & $2.679(9)$  & $2.85(8)$ & $2.62(8)$ & $2.29(8)$
& $2.81(38)$  & $2.58(8)$
\\ \hline
$10$ & $50$ & $ 40^{ 0} $ & $14.46(22)$ & $15.04(5)$ & $14.99(5)$ & $14.93(37)$ & $14.63(37)$ & $14.19(37)$
& $15.08(58)$ & $14.77(40)$
\\ \hline
$50$ & $10$ & $ 40^{ 0} $ & $37.69(46)$ & $38.28(13)$ & $38.12(12)$ & $38.29(12)$ & $37.87(12)$ & $37.37(12)$
& $38.4(10)$ & $38.9(9)$
\\ \hline
$0.5$ & $10$ & $ 40^{ 0} $ & $73.86(54)$ & $76.66(25)$ &
$76.31(25)$ & $73.67(22)$ & $73.57(22)$ & $72.58(22)$ & $-$ & $-$
\\ \hline
\end{tabular}
\end{table}

\begin{figure}[htbp]
\includegraphics*[width=4in]{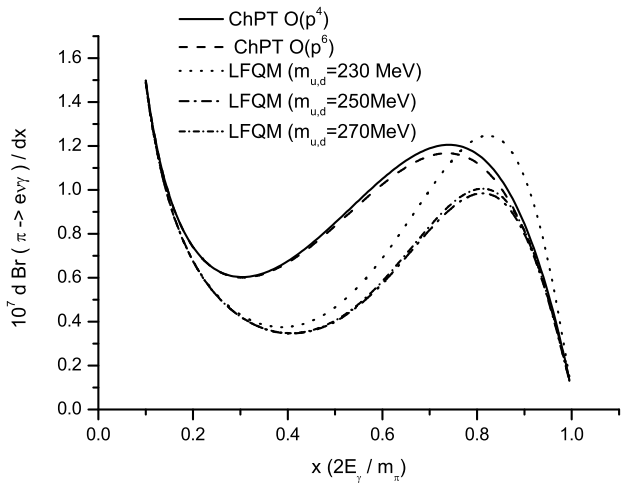}
\caption{Differential decay branching ratio
as a function of $x=2E_{\gamma}/m_{\pi}$.}
\label{dpi}
\end{figure}

In Table III, we give  the decay branching ratio of $\pi \to e \nu_e \gamma$
in various kinematic energy regions
in (a) the ChPT at $O(p^4)$, (b) the ChPT at $O(p^6)$,
(c) the LFQM, (d) the green function method~\cite{pi4} and (e) the ChPT with a large $N_C$ expansion~\cite{largenc}
as well as  the data in Ref.~\cite{pibeta2}.
Here, we have used $m_{u,d}=250$~MeV in the LFQM.
The errors in the parentheses of our results in Table III are from the decay of $\pi\to e\nu$. However, it should be noted 
that large uncertainties could arise from the various normalized coupling constants and the light quark masses in the ChPT
and LFQM, respectively.
In Fig. \ref{dpi}, we display the spectrum of the differential decay
branching ratio as a function of $x=2E_{\gamma}/m_{\pi}$ in the ChPT at both $O(p^4)$ and $O(p^6)$ and the LFQM.
 From Table~III, we see that
the results of the ChPT at $O(p^4)$ and $O(p^6)$ are higher than those of the experimental data,
which can be understood
from Table I and Eq. (\ref{48}) since
the values of $F_{V-A}(0)\equiv F_V(0) -F_A(0)$ in the ChPT
 are larger than those  fitted
in the experimental data. 
%However, our result from the LFQM  gives the closest value to the data among the theoretical expectations
%in the table.
Since the result from the LFQM agrees well with the data~\cite{pibeta2}, it could lead to a strong constraint on new physics.
We now examine the contribution to the decay from
the  tensor interaction in Eq. (\ref{tensorl}).
 From  Eqs. (\ref{tensorl}) and (\ref{tff4}), one obtains the
new tensor contribution as
\be
M_{T}&=&i e{G_F \over \sqrt{2}} \sin\theta_c
\left(F_{T}\,\epsilon^{*}_{\mu}\,q_{\nu}\,\right)
\left[\,\bar{l}\,\sigma^{\mu\nu}(1-\gamma_{5})\,\nu_{\l}\,\right]\,.
\label{ampten}
\ee
 Due to the above  tensor interaction, Eq. (\ref{43}) should be rewritten as follows:
\be
A=A_{IB}(x,y)+A_{SD}(x,y)+A_{INT}(x,y)+A_{TT}(x,y)+A_{IBT}(x,y)+A_{SDT}(x,y)\,,
\ee
where the new terms are given by
\be
A_{TT}(x,y) &=&
\frac{m_{\pi}^{4}}{ f^2_{\pi}m_{e}^{2}} |F_{T}|^{2}\,x^{2}\,\lambda (1-\lambda)\,,
  \nonumber \\
A_{IBT}(x,y) &=&
2 \frac{m_{\pi}^{2}}{f_{\pi}m_{e}} Re(F_{T})
\left( 1+r_{e}-\lambda-\frac{r_{e}}{\lambda}\right)\,,
  \nonumber \\
A_{SDT}(x,y) &=&\frac{m_{\pi}^{3}}{ f^2_{\pi}m_{e}}Re(F_{T})
(F_{V}-F_{A})\,x^{2}\,\lambda (1-\lambda) \,.
\label{tenAmp}
\ee
Integrating over x and y variables in Eq.(\ref{tenAmp}) and using the form factor
in Eq.(\ref{ttff}), we get the tensor related parts of the branching ratio
as shown
 in Table. IV.
\begin{table}[htbp]
\caption{ The tensor related parts of the  decay branching ratio for $\pi \to e \nu_e \gamma$
(in units of $10^{-8}$)
in the LFQM with the cuts of $E_{\gamma} > $ 50 MeV and
$E_{e} > $ 50 MeV.}
 \vskip 0.2in
\label{Table4}
\begin{tabular}{|c|c|c|} \hline
 TT & SDT & IBT
\\ \hline \hline
 $ 4.636 \times 10^{2}\,f_{T}^{2} $ & $9.817\times 10^{-2}\,f_{T} $
& $2.536\times 10\,f_{T} $
\\ \hline
\end{tabular}
\end{table}
To evaluate
the total branching ratio including the tensor part,
we can  combine the results in Tables II and IV.
By comparing the final result with
the experimental data of $\pi^{\pm} \to e^{\pm}\nu_{e}\gamma$,
we  extract $f_{T} = (3.48^{+8.02}_{-8.23})\times 10^{-4}$
\begin{table}[htbp]
\caption{Form factor of $f_T$ in units of $10^{-4}$.}
\vskip 0.2in
\begin{tabular}{|c|c|c|c|c|c|} 
\hline
  LFQM & \cite{pibeta2}  & \cite{tensor} & \cite{tensor2} & \cite{tensor3} &\cite{pi4}
\\ \hline \hline
  $ 3.48^{+8.02}_{-8.23}$
 & $-0.6 \pm 2.8$
& $ -56\pm 17$ 
& $ 372\pm 120 $
& $-115\pm 33$   
& $1\pm14$ 
\\ \hline
\end{tabular}
\label{Table5}
\end{table}
 shown in Table~\ref{Table5}
   in the LFQM for $m_{u,d}=250$~MeV.
In the table, we have also given other results in the literature including
the single tensor form factor fitted by PIBETA~\cite{pibeta2}. We note that
our result in the LFQM and that by PIBETA correspond to
$-1.0\times 10^{-3}<f_T<1.66\times 10^{-3}$ and $-5.2 \times 10^{-4} <f_T< 4.0 \times 10^{-4}$
  at $90\%$ C.L., respectively.

\section{Conclusions}

We have studied the momentum dependent $\pi \to \gamma$
transition form factors $F_{A,V}(p^2)$ in the ChPT and LFQM.
In particular, we have found that
$F_{A}(0)=0.0112$, $0.0102$, and $(0.151,0.0131,0.0113)$ and
$F_{V}(0)=0.0272$, $0.0272$, and $(0.0275,0.0261,0.0243)$
in (a) the ChPT $O(p^4)$,  (b) the ChPT $O(p^6)$, and  (c) the LFQM with  $m_{u,d}=(230, 250, 270)$~MeV, respectively,
at the maximal recoil of $p^2=0$,
Based on these form factors,  we have
calculated the decay branching ratio of $\pi \to e \nu_{e} \gamma$.
Explicitly, we have obtained that in the SM with the cut of
$E_{\gamma}>m_e$ and $E_{e}>$
10 MeV with the relative angle $\theta_{e\gamma}>40^o$, the decay branching ratio is
$76.66\pm0.25$, $76.31\pm0.25$ and $(73.67\pm0.22,73.57\pm0.22,72.58\pm0.22)\times 10^{-8}$
 in (a), (b) and (c), respectively, while the experimental measurement is
 $73.86\times 10^{-8}$ by the PIBETA Collaboration.
 Since our results fit well with the data, we have also derived a constraint for the tensor interaction
 %with the upper bound of the coupling 
 to be 
 $-1.0\times 10^{-3}<f_T<1.66\times 10^{-3}$ 
  at $90\%$ C.L.
 %$f_T<1.2\times 10^{-3}$ 
 %at $1\sigma$ level
in the LFQM .

\section{Acknowledgments}
We thank Professors D.A. Bryman and T. Numao for useful discussions.
This work is supported in part by the National Science Council of
R.O.C. under Grant \#s:
NSC97-2112-M-006-001-MY3 (CHC), NSC-95-2112-M-007-059-MY3 (CQG),
NSC-98-2112-M-007-008-MY3 (CQG) and NSC-97-2112-M-471-002-MY3 (CCL)
and by the Boost Program of NTHU (CQG).
% the National Center for Theoretical
%Sciences, Taiwan.


\begin{thebibliography}{99}

\bi{istra}
V. N. Bolotov et. al., Phys. Lett. {\bf B243} (1990) 308; Sov. J. Nucl. Phys. {\bf 51} (1990) 455.

\bi{pdg}
C.~Amsler {\it et al.}  [Particle Data Group],
  %``Review of particle physics,''
  Phys.\ Lett.\  B {\bf 667} (2008) 1.
  %%CITATION = PHLTA,B667,1;%%
%Particle Data Group, J. Phys. {\bf G33}, 1 (2006).

\bi{pibeta1}
E. Frlez, et. al., Phys. Rev. Lett {\bf 93} (2004) 181804.

\bi{pibeta2}
M.~Bychkov {\it et al.},
  %``New Precise Measurement of the Pion Weak Form Factors in the Pi+ -> e+ nu
  %gamma Decay,''
  Phys.\ Rev.\ Lett.\  {\bf 103} (2009) 051802.
%  [arXiv:0804.1815 [hep-ex]].
  %%CITATION = PRLTA,103,051802;%%
%M. A. Bychkov, talk given at the APS April Meeting-2007, April 14th-17th (2007).

\bi{PEN}
See the PENCollaboration Home Page at  http://pen.phys.virginia.edu/ and references therein.

\bi{Pocanic}
%\bibitem{Pocanic:2009ge}
  D.~Pocanic {\it et al.}  [PEN Collaboration],
  %``PEN: a sensitive search for non-(V-A) weak processes,''
  arXiv:0909.4360 [hep-ex]; see also
  %%CITATION = ARXIV:0909.4360;%%
 D.~Pocanic,
  %``Pion, muon decays and weak interaction symmetries,''
  arXiv:1004.4192 [hep-ex].
  %%CITATION = ARXIV:1004.4192;%%

  \bi{TRIUMF}
   D.~Bryman,
  %``Prospects for measurements of the pi $\to$ e nu branching ratio,''
  PoS {\bf KAON} (2008) 052.
  %%CITATION = POSCI,KAON,052;%%

\bi{mod2}
J. Bijnens, G. Ecker and J. Gasser, Nucl. Phys. {\bf B396} (1993) 81.

\bi{mod3}
L. Ametller, J. Bijnens, A. Braman and F. Cornet, Phys. Lett. {\bf B303} (1993) 140.

\bi{mod4}
C. Q. Geng, I. L. Ho and T. H. Wu, Nucl. Phys. {\bf B684} (2004) 281.

\bi{tensor}
A.~A.~Poblaguev,
  %``On the pi $\to$ e neutrino gamma decay sensitivity to a tensor coupling in
  %the effective quark lepton interaction,''
  Phys.\ Lett.\  B {\bf 238} (1990) 108.

 \bi{tensor1}
% \bibitem{Belyaev:1991gs}
  V.~M.~Belyaev and I.~I.~Kogan,
  %``Supersymmetry and tensor coupling in pi- $\to$ e- anti-electron-neutrino
  %gamma decay,''
  Phys.\ Lett.\  B {\bf 280}, 238 (1992);
  %%CITATION = PHLTA,B280,238;%%
  Yu.~Y.~Komachenko,
  %``Charged Higgs bosons in pi $\to$ e anti-electron-neutrino gamma decay,''
  Sov.\ J.\ Nucl.\ Phys.\  {\bf 55}, 1384 (1992)
  [Yad.\ Fiz.\  {\bf 55}, 2487 (1992)].

   \bi{tensor2}
M. V. Chizhov, Mod. Phys. Lett. {\bf A8} (1993) 2753;
%\bi{pi3}
%M. V. Chizhov,
Phys. Part. Nucl. Lett. {\bf 2} (2005) 193.

\bi{ffdef}
C. Q. Geng and S. K. Lee, Phys. Rev. {\bf D51} (1995) 99.

\bi{tensor3}
A. A. Poblaguev, Phys. Rev. {\bf D68} (2003) 054020.


\bi{pi4}
V.~Mateu and J.~Portoles,
  %``Form Factors in radiative pion decay,''
  Eur.\ Phys.\ J.\  C {\bf 52}, 325 (2007)
  [arXiv:0706.1039 [hep-ph]].
%J. C. Vicent Maten and J. Portoles, Accepted by Eur. Phys. J. C.

\bi{largenc}
 R.~Unterdorfer and H.~Pichl,
  %``On the Radiative Pion Decay,''
  Eur.\ Phys.\ J.\  C {\bf 55}, 273 (2008)
  [arXiv:0801.2482 [hep-ph]].
%R. Unterdorfer and H. Pichl, arXiv:0801.2482 [hep-ph].

\bi{LFQM}
W. M. Zhang and A. Harindranath, Phys. Rev. {\bf D48}, 4881 (1993);
%\bi{lf4}
K. G. Wilson {\em et al.},
%T. Walhout, A. Harindranath, W. M. Zhang, R. J. Perry and S. Glazek
Phys.Rev. {\bf D49}, 6720 (1994);
%\bi{lf2}
W. M. Zhang, Phys. Rev. {\bf D56} (1997) 1528.

\bi{vex1}
W. Jaus, Phys. Rev. {\bf D41}, 3394 (1990); {\bf 44}, 2851 (1991);
%\bi{vex2}
Demchuk et al., Phys. Atom. Nucl {\bf 59}, 2152 (1996).


\bi{Bryman}
D. A. Bryman {\em et al.}, Phys. Rep. {\bf 88}, 151 (1982).


\bi{Chen1}
C.~H.~Chen, C.~Q.~Geng and C.~C.~Lih,
  %``T violating muon polarization in K+ --> mu+ nu gamma,''
  Phys.\ Rev.\  D {\bf 56}, 6856 (1997).

\bi{lf1}
C.~H.~Chen, C.~Q.~Geng and C.~C.~Lih,
  Phys.\ Rev.\  D {\bf 77}, 014004 (2008);
  Int. J. Mod. Phys. {\bf A23}, 3204 (2008).

  \bi{mod1}
C. Q. Geng, C. C. Lih and C. C. Liu, Phys. Rev. {\bf D62} (2000) 034019.

\bi{lf2}
C.~Q.~Geng, C.~C.~Lih and W.~M.~Zhang,
  %``Radiative leptonic B decays in the light front model,''
  Phys.\ Rev.\  D {\bf 57}, 5697 (1998);
Phys.\ Rev.\  D {\bf 62}, 074017 (2000);
  Mod.\ Phys.\ Lett.\  A {\bf 15}, 2087 (2000);
%\bi{lf1}
C.~C.~Lih, C.~Q.~Geng and W.~M.~Zhang,
  %``Study of B(c)+ $\to$ l+ nu, gamma decays in the light front model,''
  Phys.\ Rev.\  D {\bf 59}, 114002 (1999).
  %C.~Q.~Geng, C.~C.~Lih and W.~M.~Zhang,
  %``Study of B(s,d) --> l+ l- gamma decays,''
%Wei-Min Zhang, Chin. J. Phys. {\bf 31} (1994) 717.

%\bi{ch1}
%A. Dobado, M. J. Herrero and T. N. Truong, Phys. Lett. {\bf B235}, 134 (1990).
%\bi{ch2}
%O. Strandberg, hep-ph/0302064.

\bi{kl2m}
J. F. Donoghue and B. R. Holstein, Phys. Rev. {\bf D40}, 2378 (1989);
Phys. Rev. {\bf D40}, 3700 (1989).

\bi{33}  G.~Amoros, J.~Bijnens and P.~Talavera,
  %``QCD isospin breaking in meson masses, decay constants and quark mass
  %ratios,''
  Nucl.\ Phys.\  B {\bf 602}, 87 (2001)

\bi{29}
O.~Strandberg,
  %``Determination of the anomalous chiral coefficients of order p**6,''
  arXiv:hep-ph/0302064.

\bi{25}
M.~Knecht and A.~Nyffeler,
  %``Resonance estimates of O(p**6) low-energy constants and QCD  short-distance
  %constraints,''
  Eur.\ Phys.\ J.\  C {\bf 21}, 659 (2001).

\bi{Bi2007}
J.~Bijnens,
  %``Chiral perturbation theory beyond one loop,''
  Prog.\ Part.\ Nucl.\ Phys.\  {\bf 58}, 521 (2007), and references therein.



%\bi{lf3}
%W. M. Zhang and A. Harindranath, Phys. Rev. {\bf D48} (1993) 4881.

%\bi{lf4}
%K. G. Wilson, T. Walhout, A. Harindranath, W. M. Zhang, R. J. Perry and S. Glazek
%Phys.Rev. {\bf D49} (1994) 6720-6766.


\bi{lfpa}
C.~Q.~Geng, C.~W.~Hwang, C.~C.~Lih and W.~M.~Zhang, Phys.\ Rev.\  D {\bf 64}, 114024 (2001); 
C.~H.~Chen, C.~Q.~Geng, C.~C.~Lih and C.~C.~Liu, Phys.\ Rev.\  D {\bf 75}, 074010 (2007).
 


\end{thebibliography}
\end{document}